# On Digital Currency and the Transfer of World Wealth and Technology Centers

H. J. Cai

School of Computer Science, Wuhan University

Zall Research Institute of Smart Commerce, Hubei, China

hjcai@whu.edu.cn

**Abstract.** The emergence and transfer of wealth promote the evolution of civilizations. Through the pursuit of the form of wealth valued by the members of society, the self-assertiveness demands of a society can be met and thus stimulate creativity. As means of overdrawing the future, sovereign currency and bonds have gradually become modern forms of wealth and have strongly promoted scientific and technological progress and social development. However, due to the unequal distribution of wealth, the sustainability of sovereign currency and bonds is not certain. The world has been changing rapidly since the outbreak of COVID-19, and new forms of wealth need to be constructed as an extension of the *Self* of the masses, among which digital currency may be an effective carrier of value. China is on an upward trajectory, and the complex and volatile global environment can provide an opportunity for China to focus on developing aspects of its science and technology, optimize its system of governance and strengthen its internal driving force.

## 1 Introduction

Monetary theory is an important and controversial field. Competitive monetary easing and monetization of fiscal deficit are hotly debated topics in light of the current global economic slowdown, especially in the context of the current epidemic. Modern Monetary Theory (MMT) posits a government credit-based monetary system, in which the sovereign currency is subject to the actual constraint of inflation but not to the nominal budget and in which the fiscal department, rather than the central bank, assumes the functions of realizing full employment and maintaining inflation stability.

In response to the outbreak of COVID-19, the monetary policies implemented by the United States and other Western countries have been MMT policies. For example, the United States has begun to give cash to every citizen. In just a few months, the Federal Reserve's liabilities have increased from more than $4 trillion to more than $6 trillion.

In the past two years, China has begun to deploy *new infrastructures* intensively. By the middle of April 2020, thirteen provinces, autonomous regions and municipalities had released related investment plans for key projects, and eight of them had announced a total planned investment of 33.83 trillion RMB. These plans can also be seen as the extension and innovation of MMT policy.

We have proposed elsewhere the additional implementation of a *people-centered monetary system* and the establishment of individual social security accounts transferred directly from the central bank such that the pressure of RMB appreciation can be turned into dividends to be distributed equally to citizens across the country. That property is also an aspect of MMT.

In addition to the risks posed by the global practice of modern monetary theory, China also has to deal with external pressure. Some worry that a US-led monetary union is deliberately seeking to exclude China and that the Fed's massive quantitative easing is akin to fleecing the flock. In such a tense global situation, China's position and future development require a more sober and rational analysis. We propose examining the current situation in China in terms of the relationship between the rise and fall of civilizations, the transfer of wealth and the transfer of scientific and technological centers and to respond to the pressure of MMT and *decoupling*.

**2. Self-Assertiveness Demands**

In most cases, people tend to evaluate themselves as above average within their own cognitive range, and they believe that they deserve a greater share of total wealth than they receive. This tendency is called *self-assertiveness demand*. Bloch's (1989) experimental data showed that 89% of the subjects evaluated themselves to be better than they actually were along various measures compared with their colleagues. Myers' (1993) survey showed that 90% of business managers thought their performance was better than other managers, and 86% thought they were more ethical than their colleagues. Svenson's research shows that more than 80% of people believe that they drive better than others.

Self-assertiveness demand is rigid demand and is never weaker than material needs. On one hand, people tend to affirm themselves and will drive the progress of society, but on the other hand, due to people's tendency to evaluate themselves as above average, the overall self-assertiveness demands will be greater than the social aggregate supply at current production levels. The resulting gap is a severe challenge for any administrator. Therefore, to maintain the harmony and stability of a society, the administrator must provide additional supplies to fill the gap. From the perspective of the state, there are four primary means of resource supply or wealth creation that can be utilized to fill the gap and meet the self-assertiveness demands within a society.

The first way is through learning and innovation, especially in application to institutional systems and technology levels. The innovation of systems occurs with the introduction of alternatives to existing social system, while innovation to technology is triggered by new technology and new scientific discovery. Institutional innovation enables the old relations of production to be replaced by new and more advanced ones. Learning from the outside is less resource-intensive and more explosive than independent innovation; therefore, there are many examples of latecomers taking the lead.

The second way is external acquisition, including trade with external societies to gain comparative advantages, natural territorial expansion, ancient nomadic looting, and imperialist carving up and plundering of the world. Trade is the most permanent and practical form of external acquisition. Mercantilist countries attempt to export more than they import such that the harmonious development within the country can be realized by the supplement of external wealth.

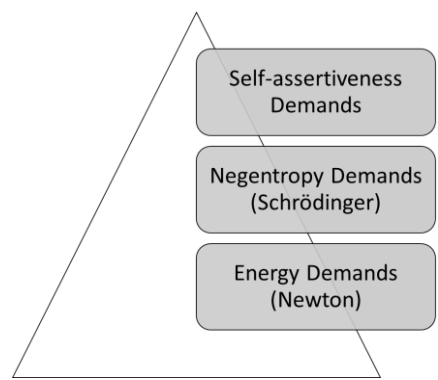

**Figure 1. Three levels of demands: when people are regarded as machines, they have energy demands; when people are regarded as ordered structures, they have negentropy demands; when people emphasize human subjectivity, they have self-assertiveness demands.**

The third way is to overdraw the future using future wealth to fill the present gap. By printing money, borrowing, and selling bonds, stocks, and other financial derivatives, we can use future resources in advance. In principle, the future is endless; therefore, the overdraft can be notably large. However, this kind of overdraft behavior is also highly variational. When the economy is stronger, people are also more optimistic, and confidence in the future is higher, meaning that the overdraft will be greater. When the economy is bad and people are pessimistic, they focus more on safeguarding their current wealth to ensure future security, and the overdraft will therefore be less. This fluctuation in confidence in the future may lead to financial market instability and financial crisis.

The last option is to start again after the crash. The transfer of wealth centers in the West in the past five hundred years has a quasi-cyclical pattern similar to the dynastic change in Chinese history. The essence of the collapse of empires or dynasties lies in the fact that the old distribution structure of wealth cannot efficiently meet the self- assertiveness demands of the whole society, and a new starting point gives people new hope. The invasion and annexation of developed countries by backward countries in recent centuries was also a feature of the change of dynasties.

**3. Transfer of Wealth and Technology Centers**

Both the East and the West have quasi-cyclical patterns of development. The development history of the West is reflected in the transfer of wealth centers between different countries, as shown in Figure 2. Earlier, Portugal and Spain co-ruled the world. Spain was responsible for the governance of the Western Hemisphere, while Portugal was responsible for the Eastern Hemisphere. Later came the period of Dutch rule followed by the rapid development and rule of the United Kingdom, and then the United States took the leading role in the development of the world.

The transfer of science and technology centers is also shown in Figure 2. The Yuasa phenomenon shows that the transfer of science and technology centers is usually not synchronized with the transfer of fortune centers in time and space but slightly lags behind, indicating that the accumulation of wealth is the first step before the emergence of innovation and creation. The center of science and technology moved from Italy to the United Kingdom, concurrent with the scientific revolution, and then to France, and then to Germany, which became famous for its chemical industry, and then to the United States. Technology centers have also tended to be unable to maintain their status as such for a long time, as the rise of a competing technology center usually had the support of sufficient accumulated wealth and could appropriate and build on the technological advances of the existing technology center. According to self-assertiveness demands, innovators in the existing technology center believed that they maintained superiority over others; therefore, the innovators spared themselves the effort of pursuing technological catch-up and innovation, making it easier for rising competitors to come to the fore.

However, in cases where the center of science and technology developed in the second and third generation, scientists began to divide into factions. In the scientific research system, obeying the rules of one's own faction will be promoted, scientific research becomes profitable, the motivations of science cease to be pure, and the nation becomes less innovative as a consequence. With the emergence of new resources, on the other hand, researchers will have purer motivations for conducting scientific research and innovation; they will find research interesting, focus on technological breakthroughs, and possibly help to form the next technology center.

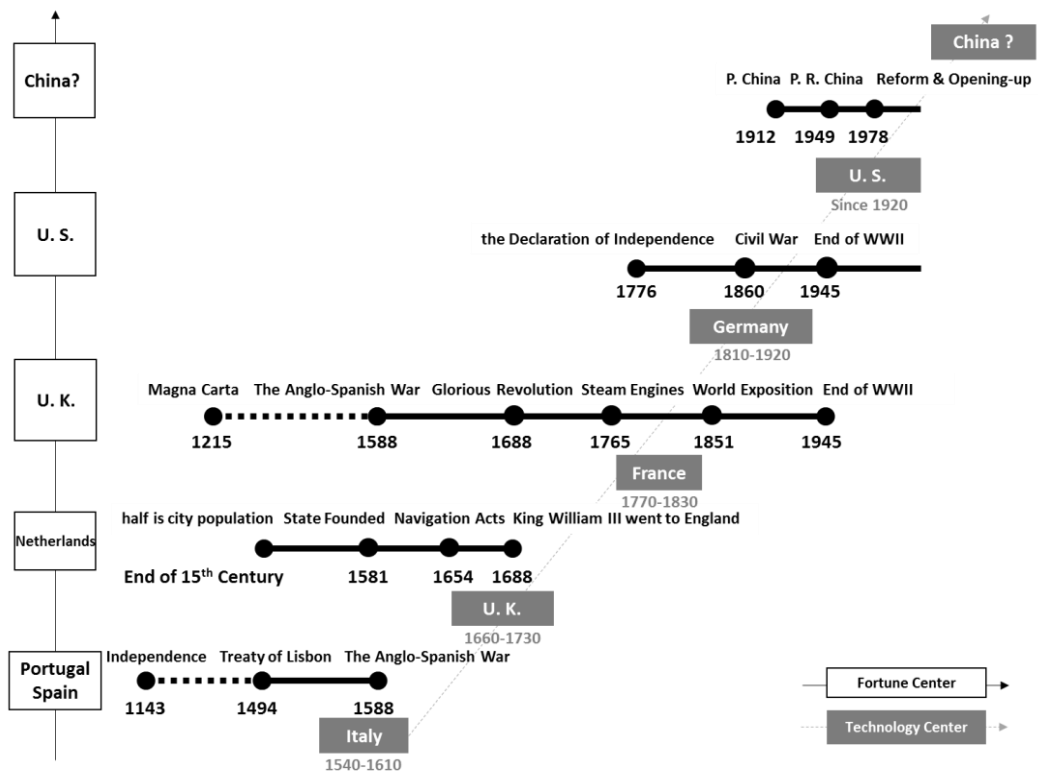

**Figure 2. Transfer of Western wealth centers and technology centers since 1500. Before the twentieth century, wealth and technology centers did not coincide in time and space, but the United States became the world's technology center in 1920 only after its industrial output (quasi-GDP) surpassed that of the United Kingdom in 1894.**

Today, the United States is both the wealth center and the technology center, its total industrial output value (quasi-GDP) having already surpassed that of the United Kingdom in 1894, and the nation having accumulated a large amount of wealth. The technology center was later transferred to the United States after 1920. Technology, of course, has further cemented America's wealth with fortunes in the form of electricity, steel, cars and oil and other commodities.

**4. Sovereign Currency and Bonds as Forms of Wealth**

Though the United States is the center of both wealth and technology, it is still not secure. The net asset appreciation rate of large American companies is approximately 10%, and the net asset growth rate of some emerging industry companies can even be as high as 14%. Even if calculated at 10%, according to Einstein's 72 Law, the assets of these companies double about every 7 years, and after 100 years of accumulation, the assets will increase by approximately 16,000 times ($2^{14}$). America's GDP grows by approximately 3% a year, or approximately 16 times in 100 years. In other words, 100 years from now, the increase in financial assets will be 1,000 times the increase in GDP, with this large gap being mainly due to the inflation of future expectations and the overdraft of the future through sovereign money and bonds.

The United States has issued a large amount of national debt during its transformation to a strong country. The outside world's expectation of its development is constantly changing. If other countries believe that the future development in the U.S. will continue to be robust, these countries will continue to pay for it. However, the United States' rolling accumulation of debt cannot last forever. Rather, it will inevitably stop growing as history progresses, possibly as a result of accidental causes. One unexpected factor is natural disasters, and another factor is external changes. For example, as long as people still believe that the United States will remain the strongest, then they will go on buying U.S. treasuries, but if there suddenly appeared an economic entity with a more promising future, then people will tend to buy the new entity's treasuries, and the United States will be fundamentally affected. This situation places the U.S. into irreconcilable conflict with emerging economies.

National debt and bonds, as means of overdrawing the future, have gradually become new forms of wealth in some countries and regions. A bond can be an asset for one group and thus a liability for another group, and these properties are largely balanced. Piketty believes that the core contradiction of capitalism is that the rate of return on capital is greater than the rate of economic growth ($r > g$), and the gap between the two must be filled by extra resources. Bonds can temporarily satisfy the annual growth demand of the middle layer (namely, interest groups) of more than 10%, which actually consume the interests of the top layer (such as the U.S. federal government) and the resources of the common people at the bottom. Many American citizens are already in a state of negative equity.

When issuing bonds, state sovereignty is highly important. Some Latin American countries' tragic situations lie in their pursuit of short-term interests, selling out state-owned assets to foreign capital, such as power plants and water plants, which are directly related to the people's livelihood. When the financial crisis took place, these governments did not have useful resources to depend on and had to print money, leading to serious currency devaluation.

Affected by the outbreak of COVID-19, the interest rate of U.S. treasury bonds continues to fall, and the U.S. has started to print money, as well. At present, since most members of the middle class are willing to continue to hold bonds as assets, there is no obvious inflation of the U.S. dollar, even as these bonds are overissued. In contrast, the U.S. dollar has recently been appreciating. However, this way of printing money still cannot solve the problem. In the longer term, the wealthiest layer will receive more of these dollars, and the ordinary people who are struggling at the bottom receive only chicken feed; as a result, the gap between the rich and the poor widens. As the challenge of sustainability will be more severe, the federal government will become more heavily indebted, the bottom will be further squeezed, and the middle layer will become dissatisfied, being unable to maintain 10% growth. The demand of the self-assertiveness of each of these parties cannot be met, which will eventually lead to social decline or even collapse.

## 5. Construction of new forms of wealth

Mr. Wu Jinglian and some other scholars believe that exchange produces wealth, which is reasonable according to the business logic of the past, and trade can thus bring large profits. However, with the intervention of Internet and Artificial Intelligence (AI) technology, commercial civilization faces new impediments and restrictions because these technologies have made commodity manufacturing transparent, and the profits are minimized. Coupled with the severe impact of COVID-19 on the global economy, it is all the more necessary to construct new forms of wealth as carriers of value such that people will be willing to pay, use and preserve them, and they therefore can serve as an extension of the *Self* of the public.

The tech elite should consider this possibility, as digital currency is a gateway. On June 18, 2019, Facebook published the Libra whitepaper, which attracted global attention, and China felt the pressure. Finance ministers from the Group of Seven (G7) are against Libra, as is the U.S. President Donald Trump, on the grounds that fiat money is on shaky ground. Although Libra maintains that it will not compete with sovereign currencies and only acts as a means of payment and a stable currency with collateral, it is bound to become a de facto competitor to fiat currencies in the context of financial instability.

Libra has adjusted its plan and is no longer linked to a basket of currencies but instead to a single currency (the U.S. dollar). This property may initially strengthen America's international position, but it will strongly threaten the U.S. dollar in the future. The U.S. dollar is endorsed by the government through future taxes, but as the world's hegemon, the U.S. has accumulated considerable debts. America's national debt exceeds $25 trillion, but its liabilities to the future are a considerably more formidable number by comparison. Such liabilities are expressed as personal assets, as well as debts owed by the state, which need to be paid continuously in the future. These liabilities are discounted to the present value of about $150 trillion. If Libra were allowed to issue digital currency, it would be nimbler than the debt-laden dollar. With Facebook having more than 2.3 billion monthly users, more than the population of any country in the world, this situation could subject the U.S. dollar to fierce competition with sovereign currencies in the near future.

A sustainable monetary system usually has a healthy ecology. However, if the gap between the rich and the poor is too large, the hierarchical structure becomes unreasonable, and the self-assertiveness demands of each layer are not satisfied for a long time, then the monetary system tends to become fragile and cannot withstand external shocks. This situation represents an unhealthy ecology. Is the ecology of the U.S. dollar system healthy and resilient to outside shocks? Libra is more similar to a special commodity than a credit currency if it is tied solely to the dollar and only used as a payment instrument, such as only allowing spending when having the assets without introducing credit.

In a normal society or a healthy ecology, credit must be used for expansion and overdrawing the future, not only paid out of the assets themselves. Many people believe that the fully collateral asset is the best solution, but we see that this situation is not the case. If everyone uses existing assets (such as gold and silver) to serve as collateral for the full specified amount, we will find that even if all social wealth were taken as collateral, it would still not be enough to fill the gap because self-assertiveness demands function unceasingly to make the requirements of the people to exceed actual social output, thereby causing challenges to sustainability, which is also the major problem with Libra lacking a credit mechanism. Therefore, credit currency is inevitable, since everyone has the motivation of pursuing wealth, and for the sake of social sustainability, the credit currency is bound to come into being. Digital currency may be an effective carrier of value as an extension of the human *Self* and even as the token of the future world.

**6. Token of the Future World**

Technologies such as AI and blockchain may challenge traditional businesses but could also provide an important basis for future forms of wealth. With the power of technology, people can have better understandings of credit characteristics; thus, the money or token of the future will be more inclined to credit, rather than actually owned assets, for collateral.

Under the recent historical trend of rapid development of technology, the essential law is still that of self-assertiveness demands such that everyone will overestimate their contribution and importance and hope to obtain more returns; this situation will still lead to credit expansion, and the over-expansion will still lead to collapse. Almost always, people's expectations will keep growing, but real wealth will not grow fast enough to catch up the expectations, and the bubble will therefore burst sooner or later.

That law leads to the first feature of the future currency, which is that it must have some inflation. For example, wage illusion is needed in companies. It is easy to increase one's salary, but it is difficult to reduce it, and this reduction will hurt the person whose salary is reduced considerably. Therefore, many companies maintain wage illusion to make the employees feel that their salaries keep rising to meet their self-assertiveness demands, but in fact, the inflation rate may offset much of the increase.

The second feature of future currency is the mechanism causing wealth to flow to the bottom. Because of self-assertiveness demands, people will seek advantages and avoid disadvantages. Wealth naturally flows to a small number of people at the top. The monetary authorities need to provide a mechanism for wealth to return to the bottom; otherwise, it will flow upward too quickly and crash more easily. Individual bankruptcy filings and food coupons in the West and secondary distribution or transfer payments in China, for example, are all ways of providing for the flow of wealth to the bottom. A good institutional system should restrain the impulse of capital as much as possible and let the social impetus be released in a slow and orderly manner.

The third characteristic of the currency of the future should be multicurrencies and should exist in an era of inexpensive money. The situation now is that Japan has retained low interest rates and negative interest rates for a long time, Europe has negative interest rates already, and the United States is moving in the same direction. Why do we emphasize multiple currencies? If all the people go up in the same value system, it is more difficult to satisfy their self-assertiveness demands. When society provides various value systems, each person may choose the most suitable value system, in which there is enough room for one's improvement. For example, many people are investing in shoe speculation, which may not be rational, but it reflects the value of collecting and hobbies, which means more than a string of numbers in the bank account and can better meet these investors' self-assertiveness demands.

People have different cognitive levels, and as AI technology advances, the differences become clearer. In the future, we will need distinguished currencies to reflect various value systems. At present, we can see such new currencies in, for example, the United States, where you can use food coupons to buy food or groceries, but you cannot use them to invest or buy luxuries. In another example, China has currently made targeted cuts in the reserve requirement ratio and industrial policies that can be implemented in different currencies in the future.

## 7. Sponsored Tokens

In the modern financial system, tokens or digital certificates can be positioned as a tool for quickly reaching consensus in a limited domain. Sovereign currency is the consensus of the whole country, representing a broad consensus that one can buy anything in the country with the fiat currency. In some countries, the government issues fiat currency, but people fail to reach a consensus, and fiat currencies can depreciate quickly. At present, many blockchain digital currencies are down to zero: that is, although the digital currency has been issued, its users are still unable to form a consensus. Digital credentials or tokens should first reach an effective consensus in a small scale and then gradually form a consensus in a larger scale.

The insurance salesman is responsible for the specific work of educating customers and upgrading their cognition according to the specific situation of each user. The insurance company can issue sales-incentive tokens according to the performance of the salesman. The salesman's income not only comes from the commission of the insurance sales but also from increases with the accumulation of the contribution of all the salesmen to the company. With the expansion of the company's scale, the value of the company will also be reflected in the growth of the whole company; therefore, the salesmen with the tokens can also share the benefits brought by the growth of the company.

That measure seems simple, but it can greatly improve the sales process and generate positive incentives for employees. Compared with companies without such a mechanism, the insurance company providing tokens will surely attract more excellent employees and thus continue to stimulate the growth of the company. In addition, token incentives can also be introduced in the design of insurance products. The design reward token and the sales reward token have different time attributes, which is more accurate and reasonable than the traditional stock and option incentives.

Using tokens in limited domain can lead to a consensus quickly, the whole process is clearly visible to all users, and the tamper-proof traceable data provide a credible witness basis. Therefore, such a mechanism of openness is beneficial to the company or industry in attracting more talent, absorbing greater external resources, and achieving greater value.

If the tokens are all placed on exchanges (that is, if the price relationship between different tokens is established), the effect is the formation of a broader consensus, which is exactly what is needed in the future. The reason why we use the work of design and sales as an example is that the completion of such work depends on subjective ability of a very high degree, as well as on excellent sales personnel, such as LI Jiaqi and Viya, whose sales can exceed those of others hundreds or even thousands of times.

Unfortunately, these advantages in the process are difficult to quantify at the beginning; therefore, we must place them in a competitive market and let everyone compete and then reach consensus. Our understanding of the many relevant factors is still in its early stages, and the fluctuations of their corresponding effects token price may be severe. In particular, highly creative cognition is still inadequate at present, tokens are needed to be issued in a limited domain to reach a consensus, and future economic behavior is actually pricing those tokens or consensus, which makes it relatively accurate.

Blockchain technology gives us a new way to reach consensus. At the beginning, it was primarily intended to reach a consensus about technology, but at present, it is gradually reflected in the community consensus, which is an inevitable direction in the future. The goal of community is ideally not to be very large, but for it to be easy to reach consensus on small issues, which will bring about changes in values. This kind of transfer of values should be encouraged. In the community, a group of people who truly believe in the value of the tokens can be gathered, and they can be allowed to find their own ways to extend themselves within the value system and meet their self-assertiveness demands.

Even if digital currencies pose a challenge to fiat currencies, we should actively respond, rather than exclude all tokens. One of the significant innovations that could be brought about in the future is to conduct Initial Token Offerings (ITO) that can be used for such purposes as governing communities and implementing blockchain reforms. Sponsors providing resources in reserve and issuing tokens with reserve can accordingly attract talents to participate in the research and development of key products or technologies and attract investors to support the project, thereby helping the organization to enhance innovation capabilities and providing new financing channels. Companies can invest in key technologies for other organizations to quickly achieve a multidimensional strategic layout, as well.

The pricing method of ITO is a composite method that combines call auctioning with the commanding price. The sponsor plays a significant role in transactions. On the one hand, the sponsor needs to hold the collateral assets as a reserve according to the number and price of issued tokens to determine the token's initial reserve rate. On the other hand, the sponsor is responsible for fulfilling users' transaction needs. In normal cases, the latest price is determined by call auctioning among all nodes involved in circulation, and then the closing price of each transaction is solved backwards. Under certain conditions, the sponsor is authorized to establish a commanding price, and the commanding price will serve as the starting price in the next round of call auctioning. The range or rule of the commanding price can be specified in the smart contract in advance. Finally, each transaction can be confirmed according to a tamper-proof, validated order record.

## 8. Summary

Most people underestimate China's role in the world in modern times and are likely to continue to underestimate its future role. In the process of China's rapid development and self-confidence building, strong public opinions and excessive behaviors are all possible. Powerful countries have all experienced similar processes. The key is to build confidence for the younger generation, reach consensus, prepare for the future, develop technology and strengthen internal motivation to embrace and lead the future. It will be easier for China to lead in the future because the West will bask in the glory it has already experienced and underestimate the changes that lie ahead.

It is not appropriate to evaluate the development potential of a country in terms of the number of patents, the number of Nobel prizes and military prowess. Japan, for example, is awarded one Nobel Prize per year on average, far exceeding China, but the prize is a lagging indicator that is largely awarded to the baby boomers of the post-World War II era. Young people born in the United States are not as active. For example, many of present technological elite are first-generation immigrants, many of the heads of big businesses are of Indian origin, and the most prominent figures in American politics are elderly, while young people are rarely observed in the political arena. Large military expenditure is also not the sign of strength. For example, rich countries often lose money in wars, because the cost to rich countries is much higher than for poor countries. If rich countries win, they will spend money like water, but if they lose, they will lose both money and status.

At present, China's GDP has reached two-thirds that of the United States, and if calculated according to the rules of purchasing power parity, China has already surpassed the United States. However, there is still room for China to grow until China's per capita GDP surpasses that of the United States. The largest advantage of China is its rising national trend.

First, China is in the stage of starting again after the collapse, and there are abundant resources to be allocated.

Second, China shares the advantage of a trade surplus, which is also a way of making external acquisitions.

Third, China has people to learn from and excellent learning ability. Even though China is characterized critically as copying others' advanced technology, it can still quickly learn and form its own technologies in many fields.

Gathering these advantages will ensure that China's development will not be crushed by setbacks. It is reasonable to believe that there will not be *decoupling* from China or *encirclement* of China by the West, because China has the most comprehensive industrial system and the largest market in the world.

While China will not voluntarily decouple, we do not need to worry too much even if decoupled passively. The United States had an early Monroe doctrine, and the United Kingdom had an early splendid isolation, taking care only of its immediate interests and not much else. China needs to learn from those experiences, but the Western route of expansion is not necessarily suitable for China. The production slowdown caused by COVID-19 may provide an opportunity for China to optimize industrial layout. For example, the inland provinces and cities can undertake coastal manufacturing industry in accordance with the principles of industrial agglomeration and supply chain optimization, thereby unleashing the vitality of technological innovation in the coastal areas.